# Twitter, human mobility, and COVID-19

Xiao Huang[1], Zhenlong Li[1*], Yuqin Jiang[1], Xiaoming Li[2], Dwayne Porter[3]

[1] *Geoinformation and Big Data Research Laboratory, Department of Geography, University of South Carolina, Columbia, SC, USA*

[2] *Department of Health Promotion, Education, and Behavior, Arnold School of Public Health, University of South Carolina, Columbia, SC, USA*

[3] *Department of Environmental Health Sciences, Arnold School of Public Health, University of South Carolina, Columbia, SC, USA*

Email[*]: zhenlong@sc.edu

**Abstract:**

The outbreak of COVID-19 is a public health pandemic that raises wide concerns worldwide, leading to serious health, economic, and social challenges. The rapid spread of the virus at a global scale highlights the need for a more harmonized, less privacy-concerning, easily accessible approach to monitoring the human mobility that has been proved to be associated with the viral transmission. In this study, we analyzed 587 million tweets worldwide to see how global collaborative efforts in reducing human mobility are reflected from the user-generated information at the global, country, and the U.S. state scale. Considering the multifaceted nature of mobility, we propose two types of distance: the single-day distance and the cross-day distance. To quantify the responsiveness in certain geographic regions, we further propose a mobility-based responsive index (MRI) that captures the overall degree of mobility changes within a time window. The results suggest that mobility patterns obtained from Twitter data are amendable to quantitatively reflect the mobility dynamics. Globally, the proposed two distances had greatly deviated from their baselines after March 11, 2020 when WHO declared COVID-19 as a pandemic. The considerably less periodicity after the declaration suggests that the protection measures have obviously affected people's travel routines. The country scale comparisons reveal the discrepancies in responsiveness, evidenced by the contrasting mobility patterns in different epidemic phases. We find that the triggers of mobility changes correspond well with the national announcements of mitigation measures, proving that Twitter-based mobility implies the effectiveness of those measures. In the U.S., the influence of the COVID-19 pandemic on mobility is distinct. However, the impacts varied substantially among states. The strong mobility recovering momentum is further fueled by the Black Lives Matter protests, potentially fostering the second wave of infections in the U.S.

# 1. Introduction

The outbreak of Coronavirus disease (COVID-19) caused by the SARS-CoV-2 virus is a public health emergency that raises wide concerns worldwide, leading to serious health, economic, and social challenges. As of June 23, 2020, there had been a total of 8,993,659 infections and 469,587 deaths globally [1], and these figures are progressively increasing every day. On March 11, 2020, the World Health Organization (WHO) reassessed the situation and officially declared COVID-19 as a pandemic, urging countries and regions worldwide to join forces [2]. Since then, major behavioral, clinical, and intervention policies (both strict and loose) have been undertaken to reduce the spread and prevent the persistence of the virus in human populations.

Initially discovered in Wuhan, China, at the end of 2019 [2], an outbreak of COVID-19 was first declared in mainland China in January 2020, before it spread out to European countries, most notably Italy, France, and the U.K, where a significant increase of cases and deaths was noticed after the outbreak in China. In the United States, the first confirmed case occurred on January 19, 2020, in Snohomish County, Washington. Shortly after, the United States has become the new epicenter of the disease as it surpassed Italy in terms of confirmed cases on March 26, 2020 [3]. As of May 31, 2020, there had been a total of 2,302,288 confirmed cases (25.6% of global cases) and 120,333 deaths (25.6% of global deaths) in the U.S. alone, according to the U.S. Centers for Disease Control and Prevention (CDC). To contain the COVID-19 pandemic, one of the non-pharmacological epidemic control measures is to reduce the transmission rate of SARS-COV-2 in the population via social distancing or other similar (self) quarantine measures [4], with the ultimate goal to reduce person-to-person interactions. Studies have found notable declines in transmission rates after the implementation of mobility-reducing policies in China, Korea, and many European countries [5-8]. Despite the success of these efforts, not all countries/regions chose to handle the pandemic in a similar manner [9, 10]. The discrepancies in policies and measures at different geographical levels urge an approach to monitoring the mobility dynamics in response to the pandemic, as mobility patterns largely indicate how people respond to the pandemic and whether policies are implemented effectively [11-13].

Since the initial outbreak of COVID-19, numerous efforts have been made, by incorporating the emerging concept of "Web 2.0" [14], "Big Data" [15], and "Citizen as Sensors" [16], to obtain timely information regarding whether people are actively reducing their exposure to COVID-19 by reducing distances traveled, and by how much. Companies like Google and Apple have released their aggregated and anonymized community mobility reports based on data collected from their mapping services (i.e., Google Maps and Apple Maps). Those reports are updated on a daily basis and can be easily downloaded. In addition, authorities started to collaborate with mobile network operators to estimate and visualize the effectiveness of control measures [11, 17], in light of the previous success of mobile phone data in assisting the modeling of the spread of other epidemics [18-20]. Shortly after the outbreak in China, mobility data from Baidu, a famous Chinese online platform, have been put into use to evaluate the effectiveness of the lockdown measure in Wuhan [5]. Leading telecommunication firms also contribute by collaborating with local authorities to estimate the efficiency of travel restrictions as well as to identify the impact of other mobility-reducing related measures [21, 22]. In the U.S., Descarte Lab (www.descarteslabs.com/mobility) has released mobility statistics derived from mobile devices, aiming to facilitate the acquisition of rapid situational awareness at the State- and County-level. City-level studies have also been conducted. For example, locational data from Cuebiq (https://www.cuebiq.com/), gathered via over 180 mobile applications, were used to monitor how social distancing guidelines are implemented on a daily basis in the city of Boston [11]. However, privacy advocates have voiced concerns on whether sharing customer data is appropriate, even in a time of crisis [23, 24]. The rapid spread of the COVID-19 at the global level highlights the need for a more harmonized, less privacy-concerning, easily accessible approach to monitoring human mobility.

The rise of social media platforms such as Twitter (twitter.com), Flickr (www.flickr.com), and Instagram (www.instagram.com), offers another possible solution to closely monitoring human mobility changes, thanks to the timely geospatial information from the enormous sensing network constituted by millions of users. The huge volume of user-generated content from social media platforms greatly facilitates the real-time or near real-time monitoring of human mobility, providing timely data of how people respond to the COVID-19 pandemic geographically, especially within different epidemic phases. The advantages of social media with respect to the aforementioned sources of digital information are that they are extensive (covering large spatial areas), easily accessible, with less privacy concern, and at very low cost [25-28]. Extracting useful information from social media is not new, as the valuable geospatial insights from social media have been explored in a wide range of fields, including hazard mitigation [29-31], evacuation monitoring [27, 32, 33], urban analytics [34-37], and public health [38, 39], to list a few. Despite the existing applications, few attempts have been made to explore the great potential of the human mobility statistics derived from social media during serious epidemic events, like the COVID-19 pandemic. Questions like whether the mobility data from social media can quantitatively reflect the collaborative effort in fighting the COVID-19 pandemic, how sensitive the mobility is during different epidemic phases, and how it corresponds to the everchanging policies in different geographical regions, deserve answers.

To answer the above questions, we focus on Twitter, a popular social media platform, and analyze more than 587 million tweets from all over the world, to see how the worldwide collaborative efforts in reducing the mobility are reflected from this user-generated information in three different scales: global scale, country scale, and U.S. state scale. We propose two types of distance, respectively termed as single-day distance and cross-day distance, to quantify different aspects of public mobility observed from Twitter. We further normalize these distances by setting up their corresponding baselines. To quantify the responsiveness in certain geographic regions within different epidemic phases, we propose a mobility-based responsive index (MRI) to capture the overall degree of mobility changes in response to the COVID-19 pandemic within a specific time window. Finally, we contextualize the mobility dynamics derived from Twitter with detailed measures from local authorities to shed light on their effectiveness. The theoretical, methodological, and contextual knowledge in this study is expected to inspire future applications of these easily accessible, less privacy-concerning, highly spatiotemporal data.

## 2. Datasets and computing environment

More than 587 million geotagged tweets posted by over 10 million twitter users are collected using the official Twitter Streaming Application Programming Interface (API), comprising a five-month period from January 1, 2020 to May 31, 2020. These tweets are stored and queried in a tweet repository managed in an in-door Hadoop cluster with 13 servers using Apache Hive and Impala. A geotagged tweet is a Twitter post with embedded geolocation in the format of exact coordinates (latitude and longitude) from the device's GPS or place names (e.g., state, county, city). While the locational accuracy of a geotagged tweet varies, depending on the settings of the account and how a user chooses to share his/her location, we exclude the tweets that are geotagged with spatial resolution lower than the city level to increase the accuracy and credibility of the mobility pattern. Following Martin et al. [40], we filter out the non-human tweets by looking at the tweet source from which application a tweet is posted. The computation of travel distance requires locational information from at least two positions. Thus, only users who post tweets on two consecutive days are included in the calculation of cross-day distance, a measure that quantifies displacement between two consecutive days (details in Section 3.1). For single-day distance, a measure that highlights the daily travel pattern, only users who post at least twice a day are included in the calculation (details in Section 3.1). Note that all distances computed in this study are Great Circle distances.

## 3. Methods

### 3.1 Single-day distance and cross-day distance

To quantify daily human mobility from collected Twitter data, we propose two different types of distance, respectively referred to as single-day distance ($D_{sd}$) and cross-day distance ($D_{cd}$). The concepts of the two distances are presented in Fig 1. In general, $D_{sd}$ represents the users' daily maximum travel distance of all locations relative to the initial location. Its calculation is confined within a single day so that users' daily travel patterns can be revealed. Different from $D_{sd}$, $D_{cd}$ measures the mean center shift between two consecutive days.

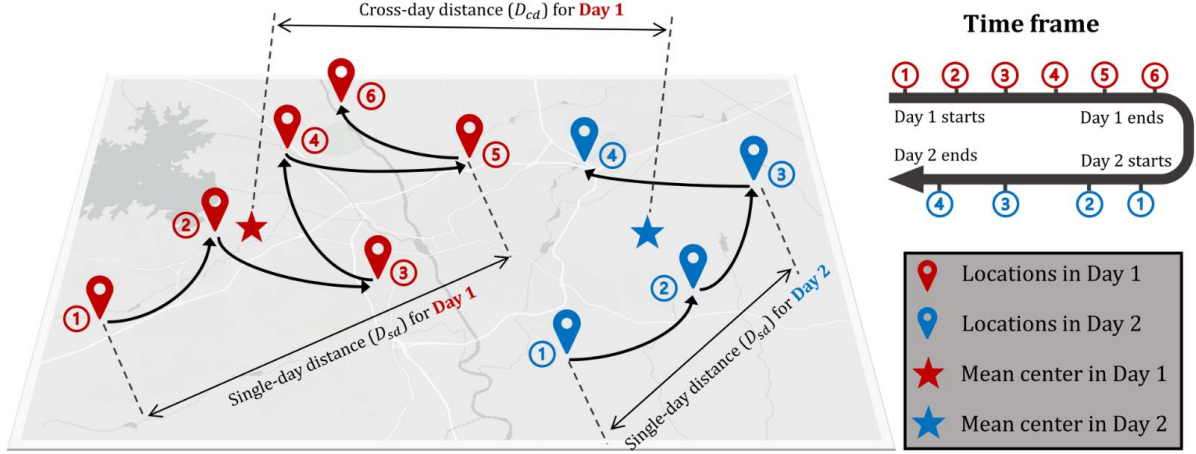

**Fig 1. Conceptualization of single-day distance ($D_{sd}$) and cross-day distance ($D_{cd}$).**

For a selected Twitter user $i$, let $P_{i,j}^m = \{P_{i,j}^1, P_{i,j}^2, \ldots, P_{i,j}^n\}$ denote the collection of locations derived from his/her tweets within a certain day $j$. Among the total of $n$ locations in day $j$, $P_{i,j}^1$ denotes the initial location and $P_{i,j}^m$ always precedes $P_{i,j}^{m+1}$ in time. To compute $D_{sd}$, a collection of location pairs (A) is first formed by coupling $P_{i,j}^m$ with the initial location $P_{i,j}^1$, i.e., A = $\{(P_{i,j}^1, P_{i,j}^2),(P_{i,j}^1, P_{i,j}^3),\ldots(P_{i,j}^1, P_{i,j}^n)\}$. The Great Circle Distance (GCD) is applied to compute the distance of each location pair within collection A. For a given location pair $(P_{i,j}^1, P_{i,j}^m)$, the GCD value between them is represented as:

$$GCD_{i,j}^{1,m} = d<\left(P_{i,j}^1, P_{i,j}^m\right)> \quad m = 1,2,3,\ldots n \tag{1}$$

where $d$ denotes the GCD operator and $GCD_{i,j}^{1,m}$ denotes GCD value of the $m$th location of user $i$ in day $j$ compared to the initial location of user $i$ in day $j$. $D_{sd}$ for user $i$ in day $j$, referred to as $D_{sd_{i,j}}$, is computed by selecting the maximum value of $GCD_{i,j}^{1,m}$, i.e., $D_{sd_{i,j}} = max\,\{GCD_{i,j}^{1,2}, GCD_{i,j}^{1,3}, \ldots, GCD_{i,j}^{1,n}\}$. To compute $D_{cd}$, for a collection of locations from user $i$ in day $j$, i.e., $\{P_{i,j}^1, P_{i,j}^2, \ldots, P_{i,j}^n\}$, a mean center ($\overline{P_{i,J}}$) is first calculated by respectively averaging the coordinates of locations in $\{P_{i,j}^1, P_{i,j}^2, \ldots, P_{i,j}^n\}$:

$$\overline{P_{i,J}} = \mu\{P_{i,j}^1, P_{i,j}^2, \ldots, P_{i,j}^n\} \tag{2}$$

where $\overline{P_{i,J}}$ denotes the mean center for user $i$ in day $j$ and $\mu$ denotes the mean center operator. $D_{cd}$ for user $i$ in day $j$, referred to as $D_{cd_{i,j}}$, is further computed by applying the Great Circle distance operator:

$$D_{cd_{i,j}} = d<\left(\overline{P_{i,J}}, \overline{P_{i,J+1}}\right)> \tag{3}$$

Intuitively, $D_{sd}$ and $D_{cd}$ represent different aspects of mobility with $D_{sd}$ measuring maximum single-day travel distance and $D_{cd}$ measuring cross-day displacement. The dynamics of $D_{sd}$ and $D_{cd}$ are expected to reflect on how the COVID-19 pandemic affects people's mobility patterns geographically, presumably indicating the regional degree of responsiveness.

### 3.2 Normalized mobility index

Inspired by the methodological design in mobility reports from Google (www.google.com/covid19/mobility) and Apple (www.apple.com/covid19/mobility), we set up baselines for $D_{sd}$ and $D_{cd}$ respectively. Unlike studies that utilize a single baseline value summarized from a fixed period of time, our mobility baselines are set for each corresponding day of a week, as a week has been widely recognized as an independent cycle in mobility [25, 41]. That is to say, we calculate a total of fourteen baseline values, seven for $D_{sd}$ and seven for $D_{cd}$, corresponding to each day of a week. For a geographical region $\mathbb{R}$ (globe, a country, or a state), let $D_{sd_j}^{\mathbb{R}}$ and $D_{cd_j}^{\mathbb{R}}$ represent the $D_{sd}$ and $D_{cd}$ of $\mathbb{R}$ in day $j$, respectively. We define that $D_{sd_j}^{\mathbb{R}}$ is the mean value of all $D_{sd_{i,j}}^{\mathbb{R}}$ in day $j$, i.e., $D_{sd_j}^{\mathbb{R}} = \frac{\sum_i D_{sd_{i,j}}^{\mathbb{R}}}{N}$, where $N$ denotes the total number of selected users in day $j$ within $\mathbb{R}$ and $P_{i,j}^1 \in \mathbb{R}$. Similarly, $D_{cd_j}^{\mathbb{R}}$ is the mean value of all $D_{cd_{i,j}}^{\mathbb{R}}$ in day $j$, i.e., $D_{cd_{i,j}}^{\mathbb{R}} = \frac{\sum_i D_{cd_{i,j}}^{\mathbb{R}}}{N}$, where $\overline{P_{i,j}} \in \mathbb{R}$. Consequently, the normalized mobility index of region $\mathbb{R}$ in day $j$ for single-day distance ($NMI_{sd_j}^{\mathbb{R}}$) and cross-day distance ($NMI_{cd_j}^{\mathbb{R}}$) are respectively defined as the ratios of $D_{sd_j}^{\mathbb{R}}$ and $D_{cd_{i,j}}^{\mathbb{R}}$ to their baseline values of a corresponding day in a week. Given their calculations, $NMI_{sd_j}^{\mathbb{R}}$ and $NMI_{cd_j}^{\mathbb{R}}$ both have a range of $[0, +\infty)$, with 1 being the critical value. When the $NMI_{sd_j}^{\mathbb{R}}$ (or $NMI_{cd_j}^{\mathbb{R}}$) is less than 1, it suggests that within region $\mathbb{R}$ in day $j$, reduced mobility is observed compared with the baseline mobility when measuring single-day distance (or cross-day distance).

### 3.3 Mobility-based responsive index

After the normalization in the previous section, a baseline of $NMI$ (i.e., $NMI = 1$) that separates patterns of increased mobility and reduced mobility is formed. Intuitively, for a time series of $NMI$ values, the size of the area under the $NMI$ baseline ($S_{AUB}$) represents the degree of positive responses (i.e., reduce in mobility) for a given period of time, while the size of the area above the $NMI$ baseline ($S_{AAB}$) indicates otherwise (Fig 2). Hypothetically, $S_{APC}$ represents a perfect scenario where mobility instantly reduced to 0 from the beginning and remains 0 until the time series ends. Conceptually, the mobility-based responsive index we propose is the ratio between the net positive response to the perfect scenario, i.e., $\frac{\sum S_{AUB} - \sum S_{AAB}}{S_{APC}}$, where $\sum S_{AUB}$ and $\sum S_{AAB}$ respectively denote the summation of areas under the curve and the summation of areas above the curve, given a specific time period.

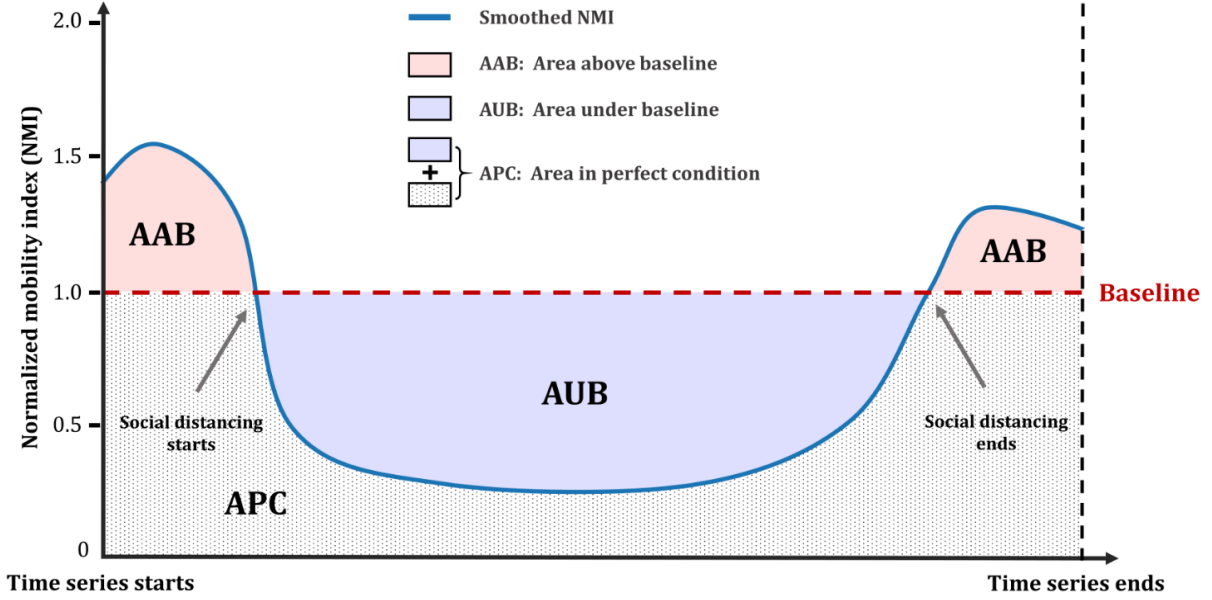

**Fig 2. Mobility-based responsive index**

To remove noises, we smooth the time series using a one-dimensional Gaussian filter ($\sigma = 2$) so that the general trend can be revealed. Further calculations regarding the size of the areas are all based on the smoothed time series. Given the different nature of the two proposed distances, we calculate their $MRI$ separately:

$$MRI_{sd} = \frac{\sum S_{AUB_{sd}} - \sum S_{AAB_{sd}}}{S_{APC}} \quad (4)$$

$$MRI_{cd} = \frac{\sum S_{AUB_{cd}} - \sum S_{AAB_{cd}}}{S_{APC}} \quad (5)$$

where $MRI_{sd}$ and $MRI_{cd}$ denote the $MRI$ with $D_{sd}$ and $D_{cd}$ being measured, respectively. We further compute an integrated $MRI$ by weighting $MRI_{sd}$ and $MRI_{cd}$ using their total sample sizes:

$$MRI = \frac{MRI_{sd} \times u_{sd} + MRI_{cd} \times u_{cd}}{u_{sd} + u_{cd}} \quad (6)$$

where $u_{sd}$ and $u_{cd}$ denote the total sample sizes used to calculate $MRI_{sd}$ and $MRI_{cd}$, respectively. The rationale of deriving an integrated $MRI$ by fusing $MRI_{sd}$ and $MRI_{cd}$ is that, despite their different calculations, they reflect human mobility from diverse perspectives, and therefore their integration serves as an overall index that better summarizes the general degree of mobility-based responsiveness geographically. The derived $MRI$ has a range of $(-\infty, 1]$. In general, the higher the value, the better responsiveness a region has, with $MRI = 1$ suggests a hypothetically perfect responsiveness. A positive $MRI$ ($MRI > 0$) suggests positive responsiveness (reduce in mobility) for a region, while a negative one suggests otherwise.

## 4. Results

### 4.1 Global scale

As most countries in the world started to aggressively respond to the COVID-19 pandemic after March 2020, we set our baselines in a temporal period from January 13 (to exclude abnormal mobility patterns due to the New Year holiday season) to February 29. Since the outbreak in China and the dramatic increase in cases in Europe, many countries have imposed and continue to impose travel bans and lockdowns [42]. As a result, both $D_{sd}$ and $D_{cd}$ have greatly deviated from their corresponding baselines, especially after March 11, when WHO declared COVID-19 as a pandemic (Fig 3a). Because both our baselines are set for the individual day in a week, their projections exhibit a clear weekly pattern. In comparison, the time series of $D_{sd}$ and $D_{cd}$, especially after the declaration of COVID-19 as a pandemic, show considerably less periodicity (Fig 3a), suggesting that the protection measures (e.g., travel restrictions, social distancing policies, stay-at-home orders) have obviously affected people's weekly routines. The visual comparison of global cross-day trajectory in February (Fig 3b) and May (Fig 3c) has also demonstrated the collaborative effort in reducing mobility to fight the pandemic, evidenced by the remarkably lower density of global trajectory in May. The gap between baselines proves the different nature of $D_{sd}$ and $D_{cd}$, well explaining our rationale of normalizing $D_{sd}$ and $D_{cd}$ separately, according to their corresponding baselines. We further observe that, throughout the entire time series, the daily value of $D_{cd}$ is considerably lower than the daily value of $D_{sd}$. This phenomenon can be explained by the existence of a large amount of Twitter users who, despite their large single-day travel distance (high $D_{sd}$ value), keep a similar daily posting routine, which leads to the no significant shift of mean centers between two consecutive days (low $D_{cd}$ value).

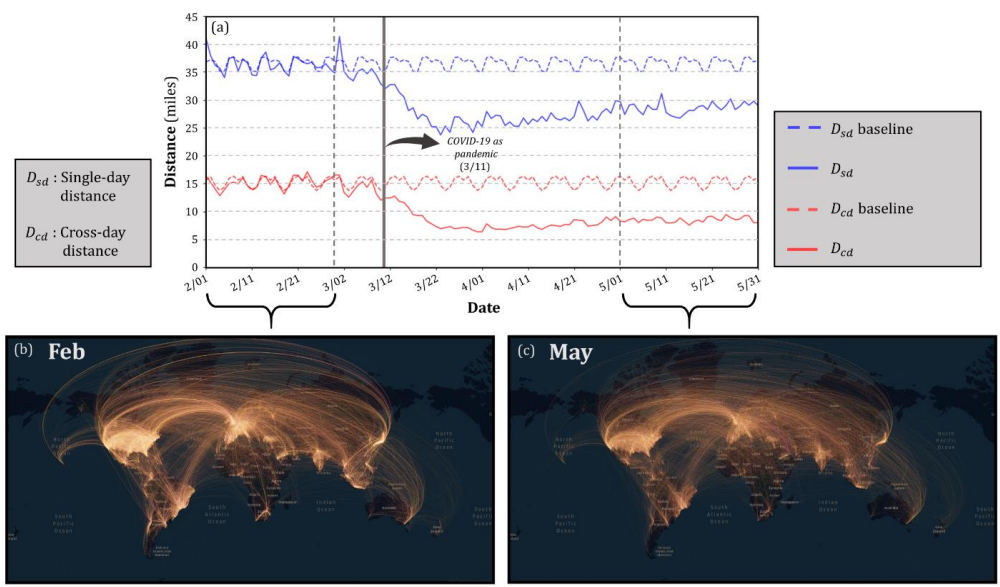

**Fig 3. Temporal distribution of global $D_{sd}$ and $D_{cd}$ in the four-month period (February, March, April, and May). (a) $D_{sd}$, $D_{cd}$ and their baselines from February to May; (b) cross-day trajectories in February; (c) cross-day trajectories in May**

We observe similar mobility dynamics when $D_{sd}$ and $D_{cd}$ are respectively normalized to $NMI_{sd}$ and $NMI_{sd}$ according to their baselines (Fig 4). Both $NMI_{sd}$ and $NMI_{sd}$ started to deviate from the baseline ($NMI = 1$) around ten days before the pandemic declaration from the WHO, suggesting that strong mobility-reducing measures had been taken before the declaration on March 11. This mobility pattern coincides with strong early travel restrictions implemented in Europe and Asia at the beginning of March [6, 43]. At the end of March, both $NMI_{sd}$ and $NMI_{cd}$ reached the bottom with the lowest $NMI_{sd} = 0.70$ and the lowest $NMI_{cd} = 0.45$, indicating that single-day distance and cross-day distance respectively reduced to 70% and 45% of the ones in the normal situation. Starting from the end of April, however, both $NMI_{sd}$ and $NMI_{cd}$ started to bounce back, and the increasing trend continued to the end of May, presumably resulting from the gradually lifted quarantine measures [44]. Compared with the hypothetically perfect scenario ($MRI = 1$) where mobility instantly halts and remains 0 throughout the time series, the overall $MRI$ for the three-month combined is 0.31, and the $MRIs$ for the March, April, and May respectively are 0.24, 0.38, and 0.32, revealing the less responsiveness in May compared with April.

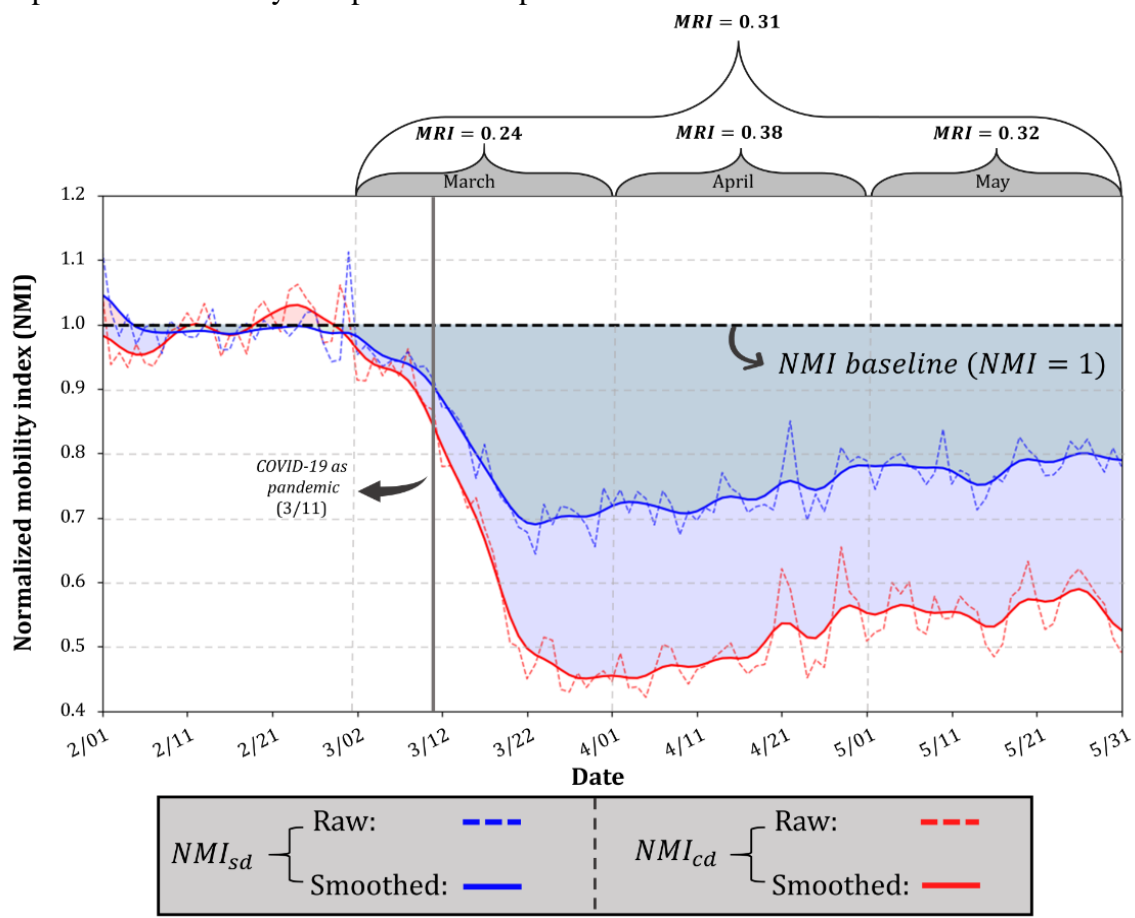

**Fig 4. Global $NMI_{sd}$ (normalized $D_{sd}$) and $NMI_{cd}$ (normalized $D_{cd}$) in the four-month period, and the monthly MRI for March, April, and May.**

## 4.2 Country scale

For country scale study, we set the mobility baseline in a period from January 13 to February 15, as some countries (e.g., Italy and South Korea) already imposed strict or voluntary mobility-reducing policies as early as in late-February. To ensure that Twitter records are sufficient enough to generate a reasonable and stable time series, we mainly target the top 20 countries with most Twitter users, according to the Digital 2020 April Global Statshot Report [45]. The selection of those countries mostly agrees with the Twitter data we collected

In general, the impact of COVID-19 pandemic on mobility derived from Twitter is obvious, as the mobility of the selected 20 countries, measured by single-day distance and cross-day distance, is mostly below the mobility baseline in March, April, and May (Fig 5), suggesting that mobility-reducing measures have been suggested and adopted in those countries. However, the country-level discrepancies in the time series of $NMI_{sd}$ and $NMI_{cd}$ can be clearly observed. The mobility in Japan started to drop in late-February (Fig 5), presumably in response to the announcement by Prime Minister Shinzo Abe on February 27 to close all Japanese elementary, junior high, and high school [46]. The further decline of mobility from early-April to late-April can be explained by the proclamation of the State of Emergency for Tokyo (April 7) and for the rest of the country (April 16) [47]. The mobility of Japan is expected to bounce back, as Japan ended the state of emergency in all of Japan On May 25 [48]. Given the limited temporal coverage of our data, however, its impact on mobility remains unknown. The mobility of the United States started to drop in mid-March when a series of statements were announced, including the declaration of COVID-19 as a pandemic by the WHO (March 11) and the declaration of National Emergency by the White House (March 13). The mobility remained consistently low in April, then gained an upward momentum in May, largely due to the gradually loosened measures [49]. A similar mobility pattern can also be observed in India, where mobility reduced following the WHO's declaration in mid-March and gradually rose in May. Mobility in Malaysia was slightly below the baseline in late-February and early-March. The sudden mobility drop appeared on March 18, which coincides with the date when the Movement Control Order (MCO) from the federal government took effect [50]. The rapid mobility reduction in Malaysia demonstrates that the MCO was effectively and efficiently executed. In Saudi Arabia, mobility started to reduce as early as March 2, when the first case was confirmed [51]. However, $NMI_{sd}$ and $NMI_{cd}$ gradually diverged as $NMI_{cd}$ remained stably low in April and May, while $NMI_{sd}$ became unstable and eventually recovered and even surpassed baseline mobility in mid-May. The divergence in trends of the two types of distances can be partially explained by the suspension of flights and mass land transport (trains, buses, and taxis) that took effect on March 21 [52]. The lack of public transit is responsible for the consistently low cross-day distance.

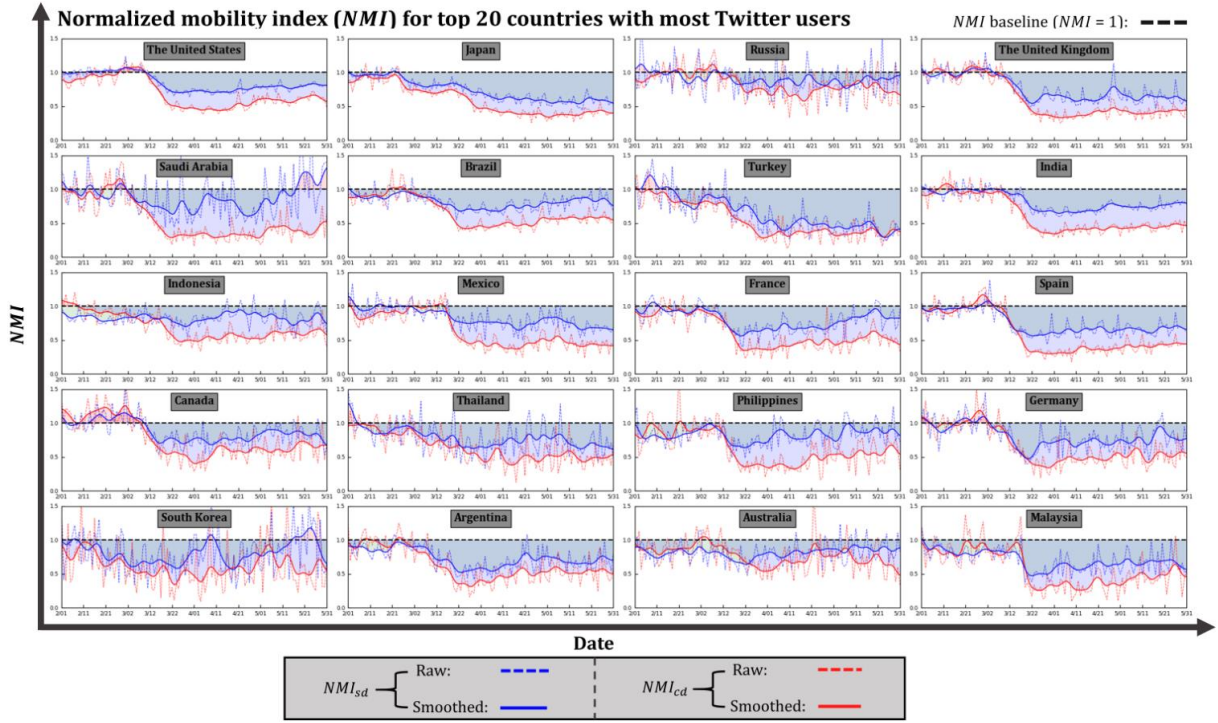

**Fig 5. Temporal distribution of $NMI_{sd}$ and $NMI_{cd}$ for the top 20 countries with most Twitter users in February, March, April, and May.**

Compared with the hypothetically perfect scenario, i.e., $MRI = 1$, Turkey has the highest three-month $MRI$ (0.50), followed by Malaysia (0.43), Spain (0.43), Japan (0.42), and the U.K. (0.41) (Table 1). Russia owns the lowest three-month $MRI$ (0.15), followed by Australia (0.21), Canada (0.26), Indonesia (0.27), and the U.S. (0.29) (Table 1). The high $MRI$ (0.42) of March in South Korea, a country that suffered from the initial spread of the epidemic in its early stage besides China, indicates that the early and strong mitigation measures were announced and implemented effectively. In light of the gradually easing situation [53], the social distancing measures in South Korea started to be lifted, evidenced by the fact that its $MRI$ decreased respectively by 0.10 and 0.12 in April and May. In the U.S., the mobility-based responsiveness in March (0.19) is among the weakest in the 20 selected countries (Table 1). In April, the $MRI$ of the U.S. reached 0.39, a net gain of 0.20 compared to the $MRI$ in March. The strong responsiveness of mobility in April is largely due to the gradually issued state-level stay-at-home orders since late-March that eventually affected at least 316 million people in at least 42 states [54]. With the lifting of orders in late-April and May, however, the U.S. showed reduced responsiveness, evidenced by its 0.1 loss in $MRI$ of May compared to April. As the U.S. has become the new COVID-19 epicenter, the reduced mobility responsiveness, along with the rocketing number of confirmed cases, deserves more attention.

**Table 1. Mobility-based Responsive index ($MRI$) for the top 20 countries with most Twitter users**

| Country names | $MRI$ | | | | | |
|---|---|---|---|---|---|---|
| | Mar | Apr | May | Three-month average | $\nabla$(Apr-Mar) | $\nabla$(May-Apr) |
| Argentina | 0.36 | 0.43 | 0.38 | 0.39 | 0.07 | -0.05 |
| Australia | 0.21 | 0.22 | 0.20 | 0.21 | 0.02 | -0.03 |
| Brazil | 0.26 | 0.39 | 0.29 | 0.31 | 0.13 | -0.10 |
| Canada | 0.21 | 0.33 | 0.24 | 0.26 | 0.13 | -0.09 |
| Germany | 0.32 | 0.40 | 0.34 | 0.35 | 0.08 | -0.06 |
| Spain | 0.32 | 0.51 | 0.46 | 0.43 | 0.19 | -0.05 |
| France | 0.31 | 0.44 | 0.32 | 0.36 | 0.12 | -0.12 |
| The United Kingdom | 0.27 | 0.49 | 0.48 | 0.41 | 0.22 | -0.01 |
| Indonesia | 0.24 | 0.29 | 0.28 | 0.27 | 0.05 | -0.01 |
| India | 0.22 | 0.45 | 0.39 | 0.36 | 0.23 | -0.06 |
| Japan | 0.25 | 0.49 | 0.52 | 0.42 | 0.24 | 0.04 |
| South Korea | 0.42 | 0.33 | 0.21 | 0.32 | -0.10 | -0.12 |
| Mexico | 0.17 | 0.42 | 0.41 | 0.33 | 0.25 | 0.00 |
| Malaysia | 0.34 | 0.53 | 0.43 | 0.43 | 0.19 | -0.10 |
| Philippines | 0.29 | 0.38 | 0.27 | 0.31 | 0.09 | -0.10 |
| Russia | 0.08 | 0.21 | 0.15 | 0.15 | 0.13 | -0.06 |
| Saudi Arabia | 0.35 | 0.46 | 0.31 | 0.38 | 0.11 | -0.15 |
| Thailand | 0.27 | 0.39 | 0.40 | 0.35 | 0.13 | 0.00 |
| Turkey | 0.30 | 0.58 | 0.62 | 0.50 | 0.29 | 0.04 |
| The United States | 0.19 | 0.39 | 0.29 | 0.29 | 0.20 | -0.10 |

Besides the countries presented in Fig 5, the temporal distribution of $NMI_{sd}$ and $NMI_{cd}$ for the other 16 countries with relatively fewer Twitter samples can be found in the Fig in S1 Fig. Information regarding the country names associated with their country codes, and the user counts for distance calculation in each country is presented in the Table in S1 Table.

### 4.2 States in the CONUS

Given that the first State of Emergency related to COVID-19 in the U.S. was declared by Washington State (WA) on February 29, while the majority of the states started to react aggressively after mid-March, we set the U.S. mobility baseline in a period from January 13 to February 29. In general, the influence of the COVID-19 pandemic on mobility is distinct, as the drop of mobility in most of the states happened in mid-March (Fig 6), potentially triggered by the events that include the pandemic declaration (March 11) and the National Emergency declaration (March 13). Although social distancing guidelines that aim to curb the spread have been suggested in the entire nation, the impacts varied substantially among states (Fig 6). Heavily hit states, e.g., NY, NJ, IL, CA, MA, and PA, generally experienced sharp mobility reduction, and their mobility remained stably low since mid-March. States with low numbers of cases, e.g., DE, MT, ME, WV, SD, and WY, despite the fluctuations in their time series, exhibited relatively marginal mobility reduction compared with heavily hit states. As the first state to announce the State of Emergency at the end of February, the mobility in WA remained close to the baseline in early March. It was not until mid-March that the mobility of WA started to noticeably decrease, which potentially indicates that the early mitigation policies in WA were not implemented effectively. The time series of mobility in states that include KS, MN, MS, AL, WV, SC, and WY, presents a bowl-shaped pattern, suggesting the strong recovery of mobility with some even bouncing beyond the baseline, due to the gradually loosened measures. In response to the COVID-19 pandemic, eight states, including AR, IA, ND, NE, SD, UT, OK, and WY, decline to impose state-wide stay-at-home orders by favoring other restrictions [54]. Without the orders, however, the aforementioned states still present considerable mobility reduction amid the pandemic, indicating the effectiveness of the federal guidelines and other mitigation approaches from the local government. Given the insufficient samples in the calculation of the baseline mobility, the mobility pattern in VT is not presented in Fig 6. The state names associated with their abbreviations and the user counts for distance calculation in each state are presented in the Table in S2 Table.

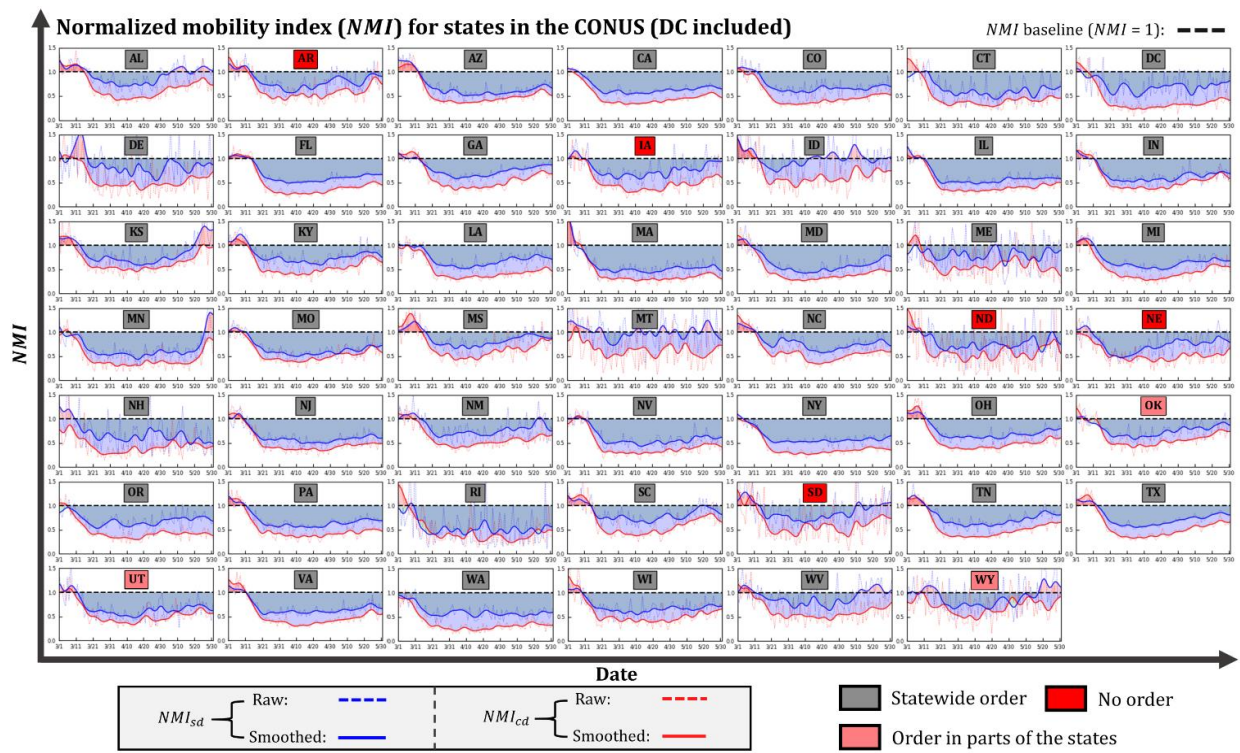

**Fig 6. Temporal distribution of $NMI_{sd}$ and $NMI_{cd}$ for states in CONUS (DC included; VT not included) in March, April, and May.**

In late-May, the risk of transmission in the U.S. was further complicated by the protests that demand justice after Mr. George Floyd died following an altercation with police. A noticeable mobility increase following the incident can be found in MN, where the incident happened (Fig 7). Carried by the existed mobility recovering momentum in mid-May, MN saw a significant increasing trend in both $NMI_{sd}$ and $NMI_{cd}$ at the end of the time series. A distinct spike can be found on May 29, 2020, when the raw $NMI_{sd}$ and raw $NMI_{cd}$ all went beyond the baseline, with the $NMI_{sd}$ (representing single-day maximum travel distance) reaching about 2.5 times than usual as a consequence of the increased activity during the protests. The divergent functionality of the smoothed $NMI$ and the raw $NMI$ is well illustrated as the former highlights the general trend while the latter is able to capture the spikes caused by disruptive events. At the time of writing, the protests have gradually spread across the U.S. and even overseas. The increase in mobility resulting from the protests deserves close monitoring, as standing in a crowd for long periods undoubtedly raises the risk of increased transmission and further worsens the situation.

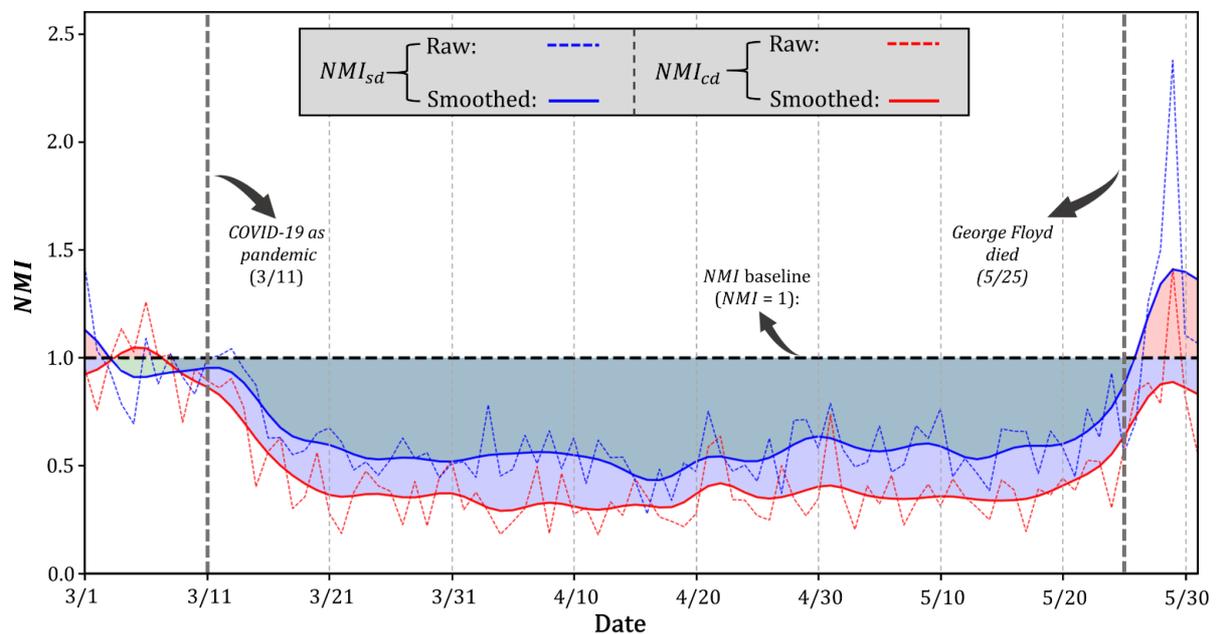

**Fig 7. Temporal distribution of $NMI_{sd}$ and $NMI_{cd}$ for Minnesota**

The monthly $MRI$ at the state level further highlights the responsiveness of each state in the three-month period. As expected, states with early spikes of cases and early strong mitigation policies tend to have a higher $MRI$ in March (Fig 8). RI (0.49) leads the $MRI$ in March, followed by WA (0.48), MA (0.45), NY (0.45), and DC (0.42) (Table 2). The high responsiveness at the early stage suggests that the mobility-reducing guidelines were implemented timely and efficiently in those states. In April, the responsiveness in almost all the states continued to strengthen (Fig 8), given the rising of cases and gradually tightened measures, except ID and MT where $MRI$ remained low and rather consistent compared to March (Table 2). From March to April, MD, AZ, and FL are the top three states with the most increase of $MRI$, respectively by 0.18, 0.17, and 0.16 (Table 2). The significant boost of mobility-based responsiveness reflects not only the severity of the situation but also the strong implementation of the mitigation measures. However, with the lifting of orders, 45 states

(except MT, NH, and WA) have shown reduced responsiveness in May compared to April (Fig 8). In light of the increasing number of cases in the U.S with no sign of slowing down (at the time of writing), the reduced mobility responsiveness can potentially foster a second wave of infections. Because of the insufficient samples in the baseline calculation, the *MRI* for VT is not presented in Fig 8 and Table 2.

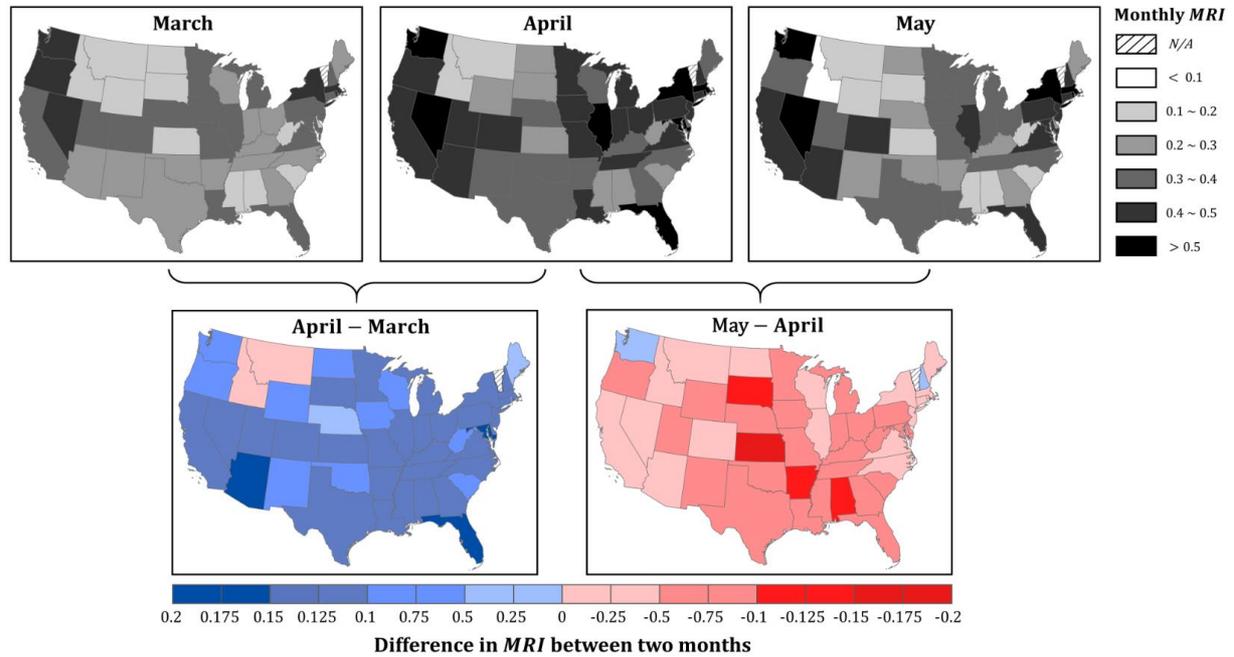

**Fig 8. Mobility-based Responsive index (MRI) for CONUS states in March, April, May, and the difference between two consecutive months.**

**Table 2. Mobility-based Responsive index (*MRI*) for states in the CONUS.**

| State abbreviations | *MRI* | | | | | |
|---|---|---|---|---|---|---|
| | Mar | Apr | May | Three-month average | ∇(Apr-Mar) | ∇(May-Apr) |
| AL | 0.14 | 0.24 | 0.11 | 0.16 | 0.10 | -0.12 |
| AR | 0.20 | 0.33 | 0.22 | 0.25 | 0.12 | -0.10 |
| AZ | 0.29 | 0.46 | 0.42 | 0.39 | 0.17 | -0.04 |
| CA | 0.36 | 0.47 | 0.42 | 0.42 | 0.10 | -0.04 |
| CO | 0.33 | 0.44 | 0.41 | 0.39 | 0.11 | -0.03 |
| CT | 0.36 | 0.50 | 0.45 | 0.44 | 0.13 | -0.04 |
| DC | 0.42 | 0.50 | 0.46 | 0.46 | 0.08 | -0.04 |
| DE | 0.19 | 0.34 | 0.27 | 0.27 | 0.15 | -0.08 |
| FL | 0.37 | 0.53 | 0.47 | 0.45 | 0.16 | -0.06 |
| GA | 0.24 | 0.36 | 0.26 | 0.29 | 0.12 | -0.10 |
| IA | 0.33 | 0.42 | 0.32 | 0.36 | 0.09 | -0.10 |
| ID | 0.12 | 0.11 | 0.09 | 0.11 | -0.02 | -0.01 |
| IL | 0.37 | 0.51 | 0.46 | 0.45 | 0.13 | -0.04 |
| IN | 0.30 | 0.42 | 0.34 | 0.35 | 0.12 | -0.07 |
| KS | 0.16 | 0.28 | 0.12 | 0.19 | 0.12 | -0.16 |
| KY | 0.23 | 0.36 | 0.27 | 0.29 | 0.13 | -0.08 |
| LA | 0.32 | 0.45 | 0.36 | 0.38 | 0.13 | -0.10 |
| MA | 0.45 | 0.58 | 0.54 | 0.53 | 0.13 | -0.04 |
| MD | 0.38 | 0.56 | 0.46 | 0.47 | 0.18 | -0.09 |
| ME | 0.29 | 0.32 | 0.29 | 0.30 | 0.03 | -0.03 |
| MI | 0.32 | 0.46 | 0.39 | 0.39 | 0.15 | -0.08 |
| MN | 0.38 | 0.49 | 0.40 | 0.42 | 0.12 | -0.10 |
| MO | 0.32 | 0.45 | 0.40 | 0.39 | 0.13 | -0.05 |
| MS | 0.14 | 0.26 | 0.16 | 0.19 | 0.12 | -0.10 |
| MT | 0.20 | 0.19 | 0.18 | 0.19 | -0.01 | 0.00 |
| NC | 0.25 | 0.37 | 0.33 | 0.32 | 0.12 | -0.04 |
| ND | 0.18 | 0.23 | 0.21 | 0.21 | 0.05 | -0.02 |
| NE | 0.32 | 0.36 | 0.29 | 0.32 | 0.04 | -0.07 |
| NH | 0.35 | 0.46 | 0.49 | 0.43 | 0.11 | 0.03 |
| NJ | 0.36 | 0.49 | 0.45 | 0.43 | 0.13 | -0.04 |
| NM | 0.25 | 0.34 | 0.26 | 0.28 | 0.09 | -0.08 |
| NV | 0.40 | 0.55 | 0.50 | 0.48 | 0.15 | -0.05 |
| NY | 0.45 | 0.56 | 0.53 | 0.51 | 0.10 | -0.02 |
| OH | 0.28 | 0.41 | 0.35 | 0.35 | 0.13 | -0.06 |
| OK | 0.26 | 0.33 | 0.24 | 0.28 | 0.07 | -0.09 |
| OR | 0.40 | 0.47 | 0.40 | 0.42 | 0.06 | -0.07 |
| PA | 0.36 | 0.48 | 0.43 | 0.42 | 0.13 | -0.06 |
| RI | 0.49 | 0.57 | 0.57 | 0.54 | 0.09 | 0.00 |
| SC | 0.15 | 0.24 | 0.17 | 0.19 | 0.09 | -0.07 |
| SD | 0.19 | 0.30 | 0.19 | 0.22 | 0.11 | -0.11 |
| TN | 0.26 | 0.41 | 0.31 | 0.33 | 0.14 | -0.09 |
| TX | 0.27 | 0.40 | 0.31 | 0.33 | 0.13 | -0.09 |
| UT | 0.33 | 0.45 | 0.39 | 0.39 | 0.12 | -0.06 |
| VA | 0.35 | 0.45 | 0.41 | 0.40 | 0.10 | -0.04 |
| WA | 0.48 | 0.54 | 0.55 | 0.52 | 0.06 | 0.01 |
| WI | 0.30 | 0.40 | 0.35 | 0.35 | 0.10 | -0.04 |
| WV | 0.18 | 0.27 | 0.17 | 0.21 | 0.09 | -0.10 |
| WY | 0.14 | 0.21 | 0.12 | 0.16 | 0.07 | -0.09 |

*Note.* VT (Vermont) is not included due to the insufficient samples in baseline calculation.

## 5. Discussion

### 5.1 Merits of social media data in gauging human mobility dynamics

The rise of social media platforms in recent years offers a potential solution to closely monitoring human mobility dynamics, given their real-time high-volume user-generated content. Public health crises like the COVID-19 pandemic uniquely highlight several merits of social media data. First, social media data are a more harmonized source compared to cellphone records from certain providers that differ geographically. Twitter, for example, has 330 million monthly active users and 500 million daily posts worldwide [55]. Its popularity allows it to serve as a valuable venue where derived mobility dynamics can be cross-compared in different regions, especially for a global epidemic event like the COVID-19. Second, social media data offer both immediacy and spatially explicit geo-information that traditional approaches like surveys and censuses are often not capable of. The rapid spread of the SARS-CoV-2 virus and the everchanging mitigation policies greatly magnify the merit of timeliness in real-time crowdsourcing platforms, including social media. Third, social media data are relatively less privacy-concerning compared to passive data-collecting approaches that include phone calls, cellular records, and smart cards. The privacy issues in the above passive methods preclude the analysis in a more spatial explicit manner, as data collected via those methods are usually de-identified and aggregated before application. Finally, social media data are easily accessible and cost-efficient. Despite the required computational resources and storage that are essential to handle the large volume and velocity of Twitter data, its data acquisition (via Twitter API) is totally free of charge. In these respects, geo-located tweets can and should be considered as a valuable proxy for human mobility, especially during times of crisis (like the COVID-19 pandemic we are facing) that usually cause dramatic mobility changes.

## 5.2 Limitations

The results of this study should be interpreted in light of several important limitations. First, the representativeness of Twitter data may not reflect the characteristics of the population as a whole in terms of socioeconomic status, age, gender, or race. Furthermore, the representativeness may vary geographically. Despite the attempts to improve the understanding of the demographics of Twitter users via profile scrutiny and tweets mining [27, 56], the intrinsic biases in Twitter samples should be considered when the results of this study are interpreted. The problem of representativeness, however, exists in all digital services. Mobility patterns derived from phone calls and cell phone applications (e.g., Google Maps and Apple Maps) also have to face the criticisms that people left behind by the "Digital Divide" [57] are significantly underrepresented.

Second, the Twitter API allows unrestricted access to only about 1% of the total records [58]. From the tweets that streamed down via the Twitter API, we only use tweets that are geotagged with spatial resolution lower than the city level. Despite the "Big Data" nature of Twitter as a data source, the available records that can be used to derive human mobility patterns are still insufficient in some regions at a temporal resolution of daily. Mobility time series computed from insufficient samples tend to have more fluctuations, making the general pattern less recognizable and less reliable. In this respect, the mobility dynamics identified in the study only account for the reaction of Twitter in response to the COVID-19 pandemic and should not be generalized to infer the mobility of the total population without caution.

Third, the less privacy-concerning nature of social media data also creates many challenges. Unlikely the passively collected data from mobile telephone records, smart cards, and wireless networks, social media data own intrinsic active nature, as users must grant permission to share their data and determine the locational accuracy of their posts, all depending on their personal settings. Thus, the two types of distances proposed in this study, single-day distance and cross-day distance, only reflect the travel behaviors that users are willing to share. This active nature protects privacy to some degree, at the same time however, dilutes the total amount of available trajectory data both spatially and temporally, potentially causing skewness in the extracted origin-destination information.

Finally, our mobility baseline, where mobility patterns from other periods are compared against, is derived from a one-month period that starts from mid-January. We further compute baselines for each corresponding day of a week by recognizing a week as an independent mobility cycle, however, without considering the monthly discrepancies that mobility patterns may present. Studies have shown that mobility may vary regularly on a monthly basis [59, 60], and the variations differ geographically due to the different cultural and societal settings. The uncertainty, resulting from the short baseline period that specifically covers late-January and February, needs to be acknowledged.

## 5.3 Future directions

Despite these limitations, we believe the strengths and valuable findings in this study outweigh the shortcomings. However, several lines of future studies are still in need. First, future work should investigate the representativeness of Twitter data by delving into the demographics of Twitter users. The mobility patterns documented in this study only reflect Twitter users' collective activities responding to the COVID-19 pandemic. Thus, the representativeness of the findings largely depends on the demographics of the local users in relation to the demographics of the local population. Another research direction is to examine the similarity and dissimilarity in mobility patterns derived from various sources (social media, phone calls, cellular records, smart cards, etc.). Although the choice of data source for studying human mobility is event dependent, a comparison study focusing on events with dramatic mobility changes leads to an overall picture and the strengths and pitfalls of each data source. The third line of research is to explore the potential in the integration of mobility indices from heterogeneous data sources. An integrated mobility index from multiple sources is expected to better reflect the multifaceted nature of human mobility, thus greatly facilitating a comprehensive mobility monitoring. Fourth, research on more extensive Twitter datasets is needed to investigate the possible improvement in mobility that can be captured. Although the hundreds of millions spatially explicit Twitter posts collected in this study are sufficient to quantitatively reflect the human mobility dynamics during the pandemic, an increasing amount of tweets are expected to generate more stable and reliable trends with fewer random fluctuations. Despite the fact that the licenses for other sample sizes, such as Gecahose (returning 10% of the public data) and Firehose (returning 100% of the public data), are costly, difficult to obtain, and requiring a demanding computational environment [61], their potential in obtaining reliable mobility dynamics at much finer spatiotemporal resolutions deserves attention. Finally, as human population movement is among the critical dimension that drives the spatial spread of COVID-19, how to leverage such twitter derived mobility information for better predicting future infectious risk of a state, county, or community warrants investigation.

## 6. Conclusion

As the whole world is now fighting the COVID-19 pandemic, the effectiveness of mobility-reducing measures (e.g., social/physical distancing) at varying scales needs rapid investigation. This article examines the reaction in social media, specifically Twitter, spatially and temporally in response to the COVID-19 pandemic as a more harmonized, less privacy-concerning, and cost-efficient approach to assessing human mobility dynamics promptly. Through analyzing around 587 million tweets worldwide, we present how our collaborative efforts in mobility reduction are reflected from this user-generated information in three different geographic scales: global scale, country scale, and U.S. state scale. To quantify various aspects of mobility from Twitter, we propose two types of distance, i.e., the single-day distance that highlights daily travel behavior and the cross-day distance that highlights the displacement between two consecutive days. To facilitate the comparison with normal situations, we further normalize these distances by separately setting up their baselines for each corresponding day of a week. We also propose a mobility-based responsive index ($MRI$) to capture the overall degree of mobility-related responsiveness of particular geographic regions in response to the COVID-19 pandemic.

The results suggest that mobility patterns obtained from Twitter data are amendable to quantitatively reflect the mobility dynamics in COVID-19 pandemic at various geographic scales. Globally, the proposed two distances measured from Twitter had greatly deviated from their baselines after March 11, 2020, when WHO declared COVID-19 as a pandemic. The considerably less periodicity after the declaration suggests that the protection measures have obviously affected people's weekly routines. The global $MRI$ reveals less responsiveness in May compared with April. At the country scale, the country-level discrepancies in responsiveness are obvious, evidenced by the contrasting mobility patterns in different epidemic phases. We further find that the triggers of mobility changes correspond well with the announcements of mitigation measures, which in return proves that Twitter-based mobility, to some degree, implies the effectiveness of those measures. At the U.S. state scale, the influence of the COVID-19 pandemic on mobility is distinct, as the drop of mobility in most of the states happened in mid-March following the National Emergency declaration on March 13. However, the impacts varied substantially among states. Heavily hit states generally experienced sharp mobility reduction while states with low numbers of cases exhibited relatively marginal mobility reduction. With orders gradually being lifted since late-April, 45 states (except MT, NH, and WA) have shown reduced responsiveness in May compared to April. The existing mobility recovering momentum is further fueled by the Black Lives Matter protests, potentially fostering the second wave of infections in the U.S. The methodological knowledge and contextual findings in this study seed future applications of the easily accessible, less privacy-concerning, highly spatiotemporal Twitter data in monitoring multi-scale mobility dynamics during disaster events.

**Funding**: The research is supported by NSF (2028791) and University of South Carolina ASPIRE program (135400-20-54176). The funders had no role in study design, data collection and analysis, decision to publish, or preparation of the manuscript.

# Supporting information

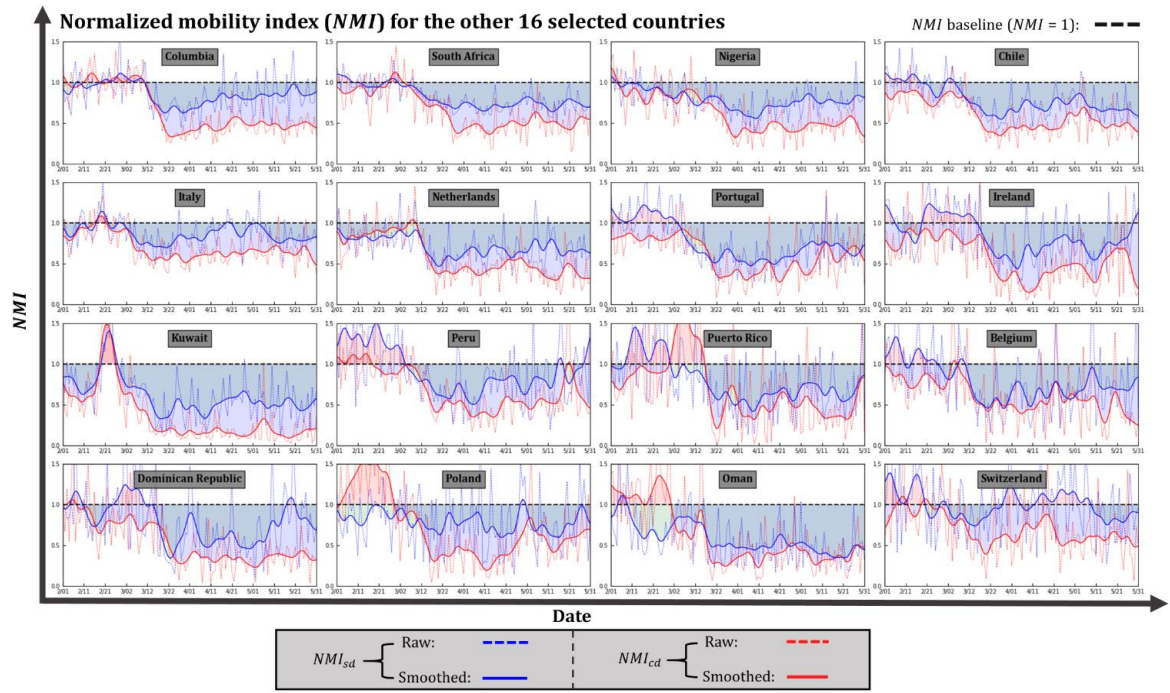

**S1 Fig. Temporal distribution of $NMI_{sd}$ and $NMI_{cd}$ for the other 16 selected countries in February, March, April, and May**

**S1 Table. Country names and user count for distance calculation in each country**

| Country names | User count for single-day distance | User count for cross-day distance |
|---|---|---|
| The United States | 17,124,880 | 16,946,890 |
| Japan | 4,216,307 | 4,204,946 |
| Russia | 360,508 | 349,698 |
| United Kingdom | 3,719,889 | 3,690,502 |
| Saudi Arabia | 1,530,731 | 1,479,899 |
| Brazil | 11,439,972 | 11,354,523 |
| Turkey | 1,139,795 | 1,031,649 |
| India | 2,419,961 | 2,148,621 |
| Indonesia | 2,950,400 | 2,809,676 |
| Mexico | 1,809,744 | 1,839,079 |
| France | 1,052,369 | 1,042,318 |
| Spain | 1,854,266 | 1,908,412 |
| Canada | 1,062,675 | 1,094,637 |
| Thailand | 696,905 | 642,702 |
| Philippines | 2,989,159 | 2,824,690 |
| Germany | 492,403 | 491,659 |
| South Korea | 164,844 | 155,097 |
| Argentina | 1,742,789 | 1,738,949 |

| | | | |
|---|---|---:|---:|
| Australia | | 518,920 | 516,405 |
| Malaysia | | 1,173,514 | 1,126,436 |
| Columbia | | 1,043,350 | 1,057,246 |
| South Africa | | 821,593 | 754,084 |
| Nigeria | | 1,024,113 | 869,689 |
| Chile | | 556,434 | 550,092 |
| Italy | | 556,168 | 573,076 |
| Netherlands | | 353,359 | 354,391 |
| Portugal | | 310,365 | 307,069 |
| Ireland | | 295,664 | 301,726 |
| Kuwait | | 298,870 | 270,495 |
| Peru | | 228,354 | 225,844 |
| Puerto Rico | | 153,077 | 153,649 |
| Belgium | | 152,627 | 151,804 |
| Dominican Republic | | 142,720 | 141,629 |
| Poland | | 138,152 | 137,128 |
| Oman | | 139,717 | 135,640 |
| Switzerland | | 80,893 | 79,807 |

**S2 Table. State abbreviations, state names, and user count for distance calculation in each state.**

| State abbreviations | State name | User count for single-day distance | User count for cross-day distance |
|---|---|---:|---:|
| AL | Alabama | 220,867 | 208,486 |
| AR | Arkansas | 94,194 | 89,937 |
| AZ | Arizona | 441,210 | 443,865 |
| CA | California | 3,331,404 | 3,355,989 |
| CO | Colorado | 273,740 | 272,968 |
| CT | Connecticut | 168,197 | 169,703 |
| DC | District of Columbia | 197,723 | 192,873 |
| DE | Delaware | 32,365 | 29,914 |
| FL | Florida | 1,257,176 | 1,191,088 |
| GA | Georgia | 644,603 | 576,796 |
| IA | Iowa | 125,301 | 123,098 |
| ID | Idaho | 51,354 | 52,955 |
| IL | Illinois | 724,546 | 727,529 |
| IN | Indiana | 328,847 | 323,840 |
| KS | Kansas | 127,859 | 129,951 |
| KY | Kentucky | 171,341 | 165,880 |
| LA | Louisiana | 411,851 | 395,070 |
| MA | Massachusetts | 391,769 | 392,283 |
| MD | Maryland | 427,486 | 415,871 |
| ME | Maine | 30,878 | 29,843 |
| MI | Michigan | 417,668 | 411,259 |
| MN | Minnesota | 221,507 | 215,945 |
| MO | Missouri | 255,015 | 251,223 |
| MS | Mississippi | 116,421 | 105,456 |
| MT | Montana | 21,563 | 21,363 |
| NC | North Carolina | 543,390 | 527,422 |
| ND | North Dakota | 22,416 | 22,144 |
| NE | Nebraska | 91,349 | 93,163 |
| NH | New Hampshire | 46,239 | 44,553 |
| NJ | New Jersey | 518,915 | 525,000 |
| NM | New Mexico | 87,255 | 92,349 |
| NV | Nevada | 316,117 | 312,085 |
| NY | New York | 1,411,974 | 1,383,876 |
| OH | Ohio | 666,406 | 658,611 |
| OK | Oklahoma | 187,695 | 185,899 |
| OR | Oregon | 218,109 | 220,195 |
| PA | Pennsylvania | 586,315 | 564,094 |
| RI | Rhode Island | 62,860 | 61,740 |
| SC | South Carolina | 191,042 | 177,017 |
| SD | South Dakota | 23,767 | 23,379 |
| TN | Tennessee | 403,233 | 397,646 |
| TX | Texas | 2,541,103 | 2,502,344 |
| UT | Utah | 131,283 | 132,005 |
| VA | Virginia | 446,301 | 428,789 |
| VT | Vermont | 17,677 | 18,289 |
| WA | Washington | 390,801 | 386,511 |
| WI | Wisconsin | 171,193 | 167,381 |
| WV | West Virginia | 56,671 | 55,580 |
| WY | Wyoming | 13,065 | 13,423 |